# Characterization of high-temperature PbTe *p-n* junctions prepared by thermal diffusion and by ion-implantation

A. V. Butenko, R. Kahatabi, E. Mogilko, R. Strul, V. Sandomirsky, Y. Schlesinger

Department of Physics, Bar-Ilan University, Ramat-Gan 52900, Israel

Z. Dashevsky, V. Kasiyan, S. Genikhov

Materials Engineering Department, Ben-Gurion University of the Negev, P.O.B. 653, Beer-Sheva 84105, Israel

**Abstract**

We describe here the characteristics of two types of high-quality PbTe *p-n*-junctions, prepared in this work: (1) by thermal diffusion of $In_4Te_3$ gas (TDJ), and (2) by ion implantation (implanted junction, IJ) of In (In-IJ) and Zn (Zn-IJ).

The results, as presented here, demonstrate the high quality of these PbTe diodes. Capacitance-voltage (*C-V*) and current-voltage (*I-V*) characteristics have been measured. The measurements were carried out over a temperature range from ~ 10 K to ~ 180 K. The latter was the highest temperature, where the diode still demonstrated rectifying properties. This maximum operating temperature is higher than any of the earlier reported results.

The saturation current density, $J_0$, in both diode types, was ~ $10^{-5}$ A/cm$^2$ at 80 K, while at 180 K $J_0$ ~ $10^{-1}$ A/cm$^2$ in TDJ and ~ 1 A/cm$^2$ in both ion-implanted junctions. At 80 K the reverse current started to increase markedly at a bias of ~ 400 mV for TDJ, and at ~ 550 mV for IJ. The ideality factor *n* was about 1.5 – 2 for both diode types at 80 K.

The analysis of the *C-V* plots shows that the junctions in both diode types are linearly graded. The analysis of the *C-V* plots allows also determining the height of the junction barrier, the concentrations and the concentration gradient of the impurities, and the temperature dependence of the static dielectric constant.

The zero-bias-resistance×area products ($R_0A_e$) at 80 K are: 850 $\Omega \cdot cm^2$ for TDJ, 250 $\Omega \cdot cm^2$ for In-IJ, and ~ 80 $\Omega \cdot cm^2$ for Zn-IJ, while at 180 K $R_0A_e$ ~ 0.38 $\Omega \cdot cm^2$ for TDJ, and ~ 0.1 $\Omega \cdot cm^2$ for IJ. The estimated detectivity is: $D^*$ ~ $10^{11}$ cm·Hz$^{1/2}$/W at $T = 80$ K, determined mainly by background radiation, while at $T = 180$ K, $D^*$ decreases to $5 \times 10^9 - 10^{10}$ cm·Hz$^{1/2}$/W, and is determined by the Johnson noise.

## 1. Introduction

We predicted recently [1, 2], on theoretical grounds, the existence of a pyroelectric effect in barrier structures of quantum paraelectrics (junction barrier pyroelectricity, JBP). The idea rests in the fact that the macroscopic dipole moment, due to the charge separation in the *p-n* junction region, depends on temperature via its dependence on the dielectric constant. The dielectric constant, in quantum paraelectrics (e.g., PbTe or $SrTiO_3$) [3], varies strongly with temperature, giving rise to a pyroelectric response. We calculated the magnitude of this effect for the quantum paraelectrics PbTe and $SrTiO_3$ [1, 2].

Thus, to test the validity of our prediction, PbTe barrier structures were chosen to investigate this effect experimentally. For this purpose, a methodology for creating high quality *p-n* junction (PNJ) barrier structures was developed. The details of the sample preparation process, resulting in outstanding diode characteristics, will be published elsewhere.

The band gap of PbTe varies from 190 meV at 0 K up to 330 meV at 300 K [4], corresponding to the IR spectral region (2 – 6 μm). A large number of publications deal with PbTe *p-n* junctions as prospective IR photovoltaic elements [4, 5]. PbTe based *p-n* junctions have been prepared in the past by a variety of methods (MBE, liquid phase epitaxy etc). In principle such diodes could be used to observe the JBP. However, for achieving a high JBP, and a better detectivity $D^*$, the diode must posses a dynamic resistance as high as possible. We also expected to find a way to extend the range of the operation temperature (the existing devices have been limited to a temperature of ~80 K).

Two types of barrier structure samples are characterized in this work. These samples differ in the method used to create the *n*-region in the original *p*-type PbTe ingot: one by $In_4Te_3$ gas diffusion (thermally diffused junctions, TDJ) and the other by indium or zinc ion implantation (implanted junctions, IJ).

We report here the results of capacitance-voltage (*C-V*) and current-voltage (*I-V*) characteristics for typical TDJ and IJ diodes in a wide temperature interval. We derived the values of the parameters essential for the interpretation of the JBP measurements, as well as for an evaluation of the photosensor qualities: the dynamical resistance at zero bias, $R_0(T)$, the zero-bias-resistance×area product, $R_0(T)A_e$, the saturation current, $J_0(T)$, the $R_0C$-time constant (*C* is the junction capacitance) and the estimated detectivity, $D^*$.

**2. Experimental**

The schematic drawing of the two types of diodes, investigated here, is presented in Fig.1.

In both cases the junction samples have been prepared on a rectangular slice of a *p*-PbTe cut from a single crystal ingot, grown by the Czochralski technique. The acceptor concentration of the as-grown crystal is $N_a \approx 10^{18}$ cm$^{-3}$. The diffusion length was ≈ 30 μm. The mobility is ≈ 14500 cm$^2$/V·s at 80 K.

For the preparation of the IJ diodes, one face of the sample has been polished and then an oxide layer has been formed by electrochemical means. Photolithography was then used to form the required pattern, as shown in Fig. 1a. An array of rectangular pits has been formed for the IJ diodes, the oxide layer serving to mask the ion beam. For the preparation of the TDJ diodes an array of mesa-structures (Fig. 1b) has been exposed to gaseous $In_4Te_3$.

The concentration of electrons, introduced by diffusion or by implantation, was $n \approx 10^{18}$ cm$^{-3}$. The junction in IJ was located at a depth of several microns under the PbTe-surface, and at about 70 μm in TDJ.

The *C-V* and *I-V* curves were measured over a temperature interval 12 – 200 K. The investigated samples were placed in a closed-cycle He gas refrigerator cryostat, in a vacuum of about $10^{-7}$ Torr. The temperature stability (using a LakeShore DRC-91CA Temperature Controller) was 0.02 K above 100 K, and about 0.002 K below 100 K.

The *I-V* characteristics of the *p-n* junctions were obtained using a 2410 Keithley SourceMeter. The applied voltage varied between –1 V and +0.5 V, measurements being taken every 0.2 mV. The positive electrode was connected to the *p* side of the diode.

A QuadTech 1920 LCR was used to measure the *C-V* characteristics. The LCR meter, used for measuring the capacitance, operates in the 4-point configuration method. The capacitance was measured using an alternating signal of 20 mV at a frequency of 1 MHz, with the applied reverse bias voltage varied between 0 V and -1 V, measurements being taken every 5 mV.

## 3. Analysis and results

### 3.1. Capacitance-voltage characteristics analysis

The *C-V* curves were measured as a function of temperature and bias voltage (*V*). In following we analyze these *C(V)* data in the framework of the standard theory of *C-V* [6]. The linearity of the $C^{-3}$ vs. *V* curves (see Fig. 2 for TDJ and Figs. 3 a, b for IJ) clearly indicates that the junctions are linearly graded. The expression (in CGSE units) for the $C^{-3}$ vs. *V* curve has the form

$$\frac{1}{C^3} = \frac{192\,\pi^2}{ea\varepsilon}V_0 + \frac{192\,\pi^2}{ea\varepsilon}V, \tag{1}$$

where *C* is the capacity per cm$^2$, the applied voltage *V*>0 for reverse voltage, and *V*<0 for forward voltage; $|-eV_0|$ is the barrier height at *V*=0 (we adopt the convention $V_0$>0); *e* is the absolute value of electron charge; *a* is the dopant concentration gradient; $\varepsilon$ is the dielectric constant. The barrier height is

$$eV_0 = E_g - F_p + F_n, \tag{2}$$

where $E_g$ is the gap, $F_p$, and $F_n$ are the Fermi-levels in *p*- and in *n*-region, respectively. $F_p$ is referenced to the upper edge of the valence band ($E_v$), while $F_n$ is referenced to the bottom of the conduction band ($E_c$):

$$F_p = E_F - E_v; \quad F_n = E_F - E_c \tag{2.1}$$

The energy gap in PbTe is known to depend on temperature [4] as

$$E_g = 0.19 + \frac{45 \times 10^{-5} T^2}{50 + T} \;[eV], \tag{3}$$

The donor concentration in a linearly graded junction is

$$N_d(x) = N_a + a\,x, \tag{4}$$

where $N_a$ = const is the acceptor concentration; $x$ is the coordinate along the p-n junction directed from p- to n-region; $x=0$ is the point, where $N_d=N_a$. It is implied that the donors and acceptors are ionized completely. The Fermi-levels are determined by net donor and acceptor concentrations at the edges of the depleted area, at $\pm W/2$,

$$W = \left[\frac{3\varepsilon}{2\pi e a}(V_0 + V)\right]^{1/3} \tag{5}$$

The net donor and acceptor concentrations are

$$N_D = N_A = aW/2 \tag{6}$$

Thus, the Fermi-levels are

$$F_n\left(\frac{W}{2}\right) = kT \ln\left(\frac{N_D}{N_c}\right); \quad F_p\left(-\frac{W}{2}\right) = -kT \ln\left(\frac{N_A}{N_v}\right) \tag{7}$$

Here, $N_{c,v}$ are the densities of states in the conduction and valence band, respectively (the factor 8 in Eq. (7.1) accounts for the 4 valleys in each band of PbTe)

$$N_{c,v} = 8\left(\frac{kTm_{n,v}}{2\pi\hbar^2}\right)^{3/2} \tag{7.1}$$

Thus, the built-in barrier, Eq. (2), is determined as the potential difference between the planes at $x = \pm W/2$.

It should be noticed that Eq. (7) implies that the charge carrier gases are not degenerate ($F_p>0$, $F_n<0$). In the general case, one should express the carrier concentrations by the corresponding Fermi integrals.

The input values in Eqs (1) – (7.1) are: the experimentally measured function $C(V,T)$; the known function $E_g(T)$, and the known densities of state $N_{c,v}$.

The procedure of deriving the diode parameters can be outlined as follows:

(*a*) The straight line, $C^{-3}$ vs. $V$, Eq. (1), provides two temperature dependent functions, namely, the slope (*M*), and the intersection (*B*) of the straight line with the ordinate axis ($C^{-3}$).

$$C^{-3} = B + M \cdot V;$$
$$M(T) = \frac{192\pi^2}{ea\varepsilon^2}; \quad B(T) = \frac{192\pi^2}{ea\varepsilon^2} V_0 \tag{8}$$

From here, $V_0$ is

$$V_0(T) = B(T)/M(T) \tag{9}$$

(*b*) Eq. (8) determines temperature dependence of the product $a \cdot \varepsilon^2$ only, but not the values $a$ and $\varepsilon$ separately.

$$a \cdot \varepsilon^2 = \frac{192\pi^2}{e \cdot M(T)} \tag{10}$$

For semiconductors, such as Si, Ge or GaAs, the dielectric constant is well known from other experiments, so that $a$ can be determined from Eq. (10). For PbTe that is not so. There exists a few works only, where $\varepsilon(T)$ was measured in a wide temperature region [4]. According to these works, $\varepsilon$ strongly increases with decrease of temperature, and depends also on the level of doping. Thus, $\varepsilon(T)$ must be determined specifically for a given sample of PbTe.

The problem of deriving independently both unknowns, $a$ and $\varepsilon(T)$, from a single equation (Eq. (10)), presents a non-trivial task. This was solved here by a semi-empirical self-consistent procedure as described in the following:

(*1*) The temperature dependence of the static dielectric constant of a ferroelectric of $ABO_3$-type is described by the Barrett's formula (BF) [7, 8]

$$\varepsilon_B(T) = \frac{C}{T_1 \cdot \coth\left(\frac{T_1}{T}\right) - T_0}, \quad (11)$$

where $T_0$ is the Curie-Weiss temperature, $T_1$ is the characteristic crossover temperature (from quantum-mechanical to classical region) and $C$ is the Curie constant. For quantum paraelectrics $T_0 < 0$. This formula describes also the experimental data for the quantum paraelectric SrTiO$_3$ [3]. *We postulate that the PbTe dielectric constant can be also described by* Eq. (11).

Thus, there are four unknown constants: $a$, $C$, $T_1$, and $T_0$. Hence, it is necessary to have two more equations in addition to Eqs. (10) and (11).

(*2*) It has been stated earlier [9], based on *experimental* evidence, that $\varepsilon$ of PbTe, in the high temperature region ($T > 200$ K), can be expressed by

$$\varepsilon(T) = \frac{1.6 \times 10^5}{T + 100} \quad (12)$$

We will consider this relationship as the third equation, valid in the region 200 K $< T <$ 300 K. Since $a$ does not depend on temperature, Eqs. (12) and (10) allow finding an initial guess value of $a$.

(*3*) The fourth relationship is obtained as follows. Differentiating both sides of Eq. (8) gives

$$\frac{dM}{dT} = -\frac{2 \times 192 \pi^2}{ea\varepsilon^3} \frac{d\varepsilon}{dT} \quad (13)$$

Since, we have postulated that $\varepsilon(T) = \varepsilon_B(T)$, Eq. (8) provides the required fourth relationship

$$-\frac{1}{2M}\cdot\frac{dM}{dT} = \frac{d\varepsilon_B}{\varepsilon_B \cdot dT} \qquad (14)$$

To get the initial guess values of the unknowns, we use the high- and the low-temperature limits of Eqs. (11) and (14). The high-temperature limit ($T_1/T \rightarrow 0$) of Eq. (11) gives:

$$\varepsilon_B(T) = \frac{C}{T - T_0} \qquad (15)$$

The low-temperature limit ($T_1/T \rightarrow \infty$) of Eq. (14) gives:

$$\ln\left(-\frac{T^2}{2\cdot M}\cdot\frac{dM}{dT}\right) = \ln\left[\frac{4T_1^2}{(T_1 - T_0)}\right] - \frac{2T_1}{T} \qquad (16)$$

The comparison of (15) and (12) provides the guess values of $C$ and $T_0$. The plot of the left hand side of Eq. (16) vs. $1/T$ provides the guess values of $T_1$ and $T_0$.

Starting with these initial value guesses, the iterative refining of the parameters $a$, $C$, $T_1$ and $T_0$, resulted in an very good fit of the dielectric constant by the Barrett's formula. The results, for both diodes, will be presented below.

(*4*) Knowing $\varepsilon(T)$, Eqs. (5) - (7) give $W$, $N_A$, $N_D$, $F_n$ and $F_p$. Thus, all the physical parameters determining the shape of the *C-V* curve, can be determined.

3.1.1 *The C-V characteristics of the* TDJ *diodes*

The characteristics properties of the TDJ diodes are presented in Fig. 4 (the dielectric constant $\varepsilon$), Fig. 5 (the barrier height, $V_0$), Fig. 6 (the dopant concentration gradient *a*), Fig. 7 (the depletion layer width $W$ and the dopant concentration at the depletion layer edge $N_D$), and Fig. 8 (the Fermi level energies $F_n$, $F_p$). These figures lead to the following conclusions:

(*i*) The experimental value of *a* depends only weakly on temperature. This validates the procedure determining $\varepsilon(T)$ and *a*.

(*ii*) The *p-n* junction width, *W*, increases with decreasing temperature, due to an increase of both $V_0$ and $\varepsilon(T)$.

(*iii*) The values of $N_D = N_A = a \cdot W/2$, by definition, are the concentrations of the dopants at the boundaries of the *p-n* junction depleted region, and determine the barrier-height and the position of the Fermi levels. In the linearly graded PNJ these boundaries, by no means, coincide with the "firm", fixed, doped region boundaries (where the dopant concentration reaches a constant value). The "firm" width of the linearly graded *p-n* junction is essentially wider, then the width of the depleted area. This fact is usually not emphasized in standard texts [6], as, for most of the common semiconductors, *W* depends only weakly on temperature. This is not the case in PbTe, since both $V_0$ and $\varepsilon$ vary significantly with temperature. *W* expands or contracts with changing temperature, and the width of the PNJ varies, correspondingly. It is clear that the varying $W(T)$ scans the "firm" profile of doping. As the donor and acceptor concentrations grow with increasing depth in the corresponding (*n*- and *p*-type) regions, the interpretation of of $N_D(T)$ in Fig. 7 becomes clear.

3.1.2 *The C-V characteristics of the* IJ *diodes*

The *C-V* plots are shown in Fig. 3 a, b. The linearity of the $C^{-3}$ vs. *V* graph expresses the fact that the IJ diodes are linearly graded as well. The analysis of experimental curves has been described in Section 3.1. The characteristic properties derived from the *C-V* results are presented in Fig. 4 (the dielectric constant $\varepsilon$), Fig. 5 (the barrier height $V_0$), Fig. 6 (the gradient of the dopant concentration *a*), Fig. 7 (the width of the depletion region *W* and the donor and acceptor concentrations $N_D$, $N_A$), and Fig. 8 (the Fermi level energies $F_n$, $F_p$).

The gradient of the dopant concentration *a* in the case of In-IJ, ~ 4.5×10$^{23}$ cm$^{-4}$, is larger then that of TDJ, ~ 2×10$^{21}$ cm$^{-4}$, by more than two orders of magnitude. Accordingly, the width of the depletion region in In-IJ, ~ 0.1 μm, is smaller by about a factor of 7 than that of TDJ, ~ 0.7 μm. For Zn-IJ the value of *a*, and *W* are comparable with those of TDJ.

The barrier height $V_0$, as shown in Fig. 5, can be determined by two independent ways: either directly from the *C-V* characteristics using Eq. (9), or by calculation according to the definition given by Eq. (2), and using the experimentally determined values of the Fermi levels. These are two independent findings of $V_0$. For TDJ the agreement between these two ways of determining $V_0$ is reasonable, but not in the case of both IJ.

According to Fig. 7, the dopant concentration in In-IJ at *W/2* is markedly higher than in TDJ and in Zn-IJ. Respectively, Fig.8 demonstrates that TDJ and Zn-IJ are non-degenerate at all temperatures ($F_n < 0$, $F_p > 0$), while In-IJ degenerates at low temperatures.

3.2. Current-voltage characteristics analysis

The experimental *I-V* curves have been fitted by Shockley's formula

$$J = J_0 \left[ \exp\left(\frac{eV}{n \cdot kT}\right) - 1 \right], \tag{17}$$

where $J_0$ is the saturation current and $n$ is the ideality factor.

The fitting parameters $J_0$ and $n$ have been determined by plotting the logarithm of the dynamical resistance $R(V) = (\partial J/\partial V)^{-1}$, calculated numerically from the *I-V* curves, as a function of the bias voltage [10]:

$$\ln\left(\frac{dJ}{dV}\right) = \ln\left(\frac{J_0 e}{n\, k_B T}\right) + \frac{V}{n\, k_B T} \tag{17.1}$$

We found this procedure much more effective, then the fitting the *I-V* directly.

3.2.1 *The I-V characteristics of the* TDJ *diodes*

Typical experimental *I-V* curves of the TDJ diode at different temperatures are presented in Fig. 9. Typical examples of the quality of fitting are shown in Fig. 10.

The plot of log$R_0$ vs. 1/$T$ ($R_0 = R_{V=0}$) is presented in Fig. 11. We observe two temperature regions with different activation energies $E_a$. A high-temperature region ($T >$ 50 K) with $E_a = 105$ meV and a pre-exponential factor of 0.091 $\Omega$, and a low temperature region ($T \lesssim 50$ K) with $E_a = 0.22 - 0.3$ meV with a pre-exponential factor of ~ 12 M$\Omega$.

The temperature dependence of the derived ideality factor $n$ and of the saturation current densities $J_0$ are shown in Figs. 12a and 12b. The $J_0$ vs. 1/$T$ plot shows again evidence of two temperature regions with different activation energies. We found in TDJ a high temperature activation energy of 135 - 146 meV, and a pre-exponential factor of $1.08 \times 10^3$ A/cm$^2$, and a low temperature activation energy of 8.58 meV and a pre-exponential factor $8 \times 10^{-8}$ A/cm$^2$.

3.2.2 *The I-V characteristics of the* IJ *diodes*

Typical experimental *I-V* curves of the IJ diodes, at different temperatures, are presented in Fig. 13. Experimental measurements and analysis are identical to that described above in the previous Section (3.2.1).

The temperature dependence of *n* and $J_0$, obtained from the fitted *I-V* plots, are given in Fig. 12. The temperature dependence of $R_0(1/T)$ is shown in Fig. 11.

We observe again, for In-IJ, two temperature regions with different activation energies. In the low temperature region (*T* < 50 K) $E_a \approx 0.22$ meV with a pre-exponential factor ~ 2 MΩ, and in the high-temperature region (*T* > 50 K) $E_a = 105$ meV and a pre-exponential factor of 2.1 Ω. For the saturation current density (Fig. 12b) the high temperature activation energy is 135 – 146 meV, with a pre-exponential factor of $8.86 \times 10^2$ A/cm$^2$; and the low temperature activation energy of 2.45 meV, with a pre-exponential factor of $4 \times 10^{-9}$ A/cm$^2$.

The high-temperature activation energy of the resistance $R_0$, for both types of diodes, is 105 meV. Both types of diodes exhibit also similar high-temperature activation energies (135 – 145 meV) of the saturation current.

There are two main contribution to the diode current [6] – the diffusion current ($J_d$) and recombination current ($J_r$). According to the common theory $\ln J_d \propto E_g(0 \text{ K})$, and $\ln J_r \propto E_g(0 \text{ K})/2$. Thus, the recombination current contribution is essential. The ideality coefficient of TDJ in the high-temperature region, $n \approx 2$ points also to this fact. However, for In-IJ the value of *n* is much larger than 2 in a sizeable part of the high-temperature region.

The low-temperature, small activation energy, is connected, perhaps, with a tunneling process. However, a direct band-to-band tunneling is highly improbable for such broad depletion regions (Fig. 7). Therefore, presumably, the band-to-band tunneling proceedes via intermediate localized states in the gap. The following facts support this hypothesis:

(1) The low-temperature transport with very small activation energy; (2) the region with small activation energy is more distinct for the narrower IJ, than for the wider TDJ (Fig. 11); (3) the resistance of IJ in this temperature region is markedly smaller, than that of TDJ (Fig. 11); (4) in this temperature region In-IJ becomes degenerated (Fig. 8), the edge of the valence band in the *p*-region being above the edge of the conduction band in the *n*-region. That must promote the tunneling. The TDJ and Zn-IJ are close to degeneration (Fig. 8).

We draw attention also to the following experimental finding: as seen in Fig. 9a, for In-IJ, the reverse current behaves anomalously, increasing with lowering temperature.

The resistance of the Zn-IJ drops weakly with lowering temperature (Fig. 12b). This is presumably connected with the decrease of the resistance of the doped edges of the *p-n* junction, due to the increased carrier mobility, at low temperature, in these regions.

### 4. Photodetector characteristics

The $R_0A$ values obtained in this work, allow to estimate the specific detectivity ($D^*$) of these diodes.

The specific detectivity of a photodiode is defined as [11]

$$D^*_\lambda = \frac{R_\lambda}{\left(\dfrac{4kT}{R_0 A_e} + 2e^2 \eta Q_B\right)^{1/2}};$$

$$R_\lambda = \eta e \lambda / hc; \quad Q_B(v_c, T) = \int_{v_c}^{\infty} J(v)\, dv \qquad (19)$$

$$J(v) = \frac{8\pi v^2}{c^2} \cdot \frac{1}{\exp(hv/kT) - 1}$$

Here, $R_\lambda$ is the current mode responsivity at a wavelength $\lambda$; $\eta$ is the quantum efficiency, $v_c$ or $\lambda_c$ are the cut-off frequency and wavelength, respectively. For the black body radiation at 300 K typical values are $\eta \approx 0.5$, and $R_\lambda$ ($\lambda_c$=4μm) $\approx$ 1.6 A/W. These values will be used for our estimates. $J(v)$ is the flux density per unit frequency interval. $R_0 A_e$ is the zero-bias-resistance×area product. Their values are given in the Table 1 for several temperatures. The detectivity consists of two contributions. The first term in the denominator is Johnson noise, the second is the background-induced shot noise. The values of $D^*$ for 12.6 K and 80 K are determined by the background radiation only. At 140 K both contributions are comparable. The column of $D^*_J$ is the detectivity, limited only by the Johnson noise.

Zogg et al [12] have made significant progress in the preparation of high-quality PbTe *p-n* junctions (see also Ref. [5]). They characterized the epitaxial PbTe *p-n* diodes grown on Si substrate with CaF$_2$ buffer layer. Their value of $R_0A$(80 K) $\approx$ 20 Ω·cm$^2$ (Fig. 4 in Ref. [12]) is more than an order of magnitude smaller than the values given in the Table 1.

Recently, Barros et al [13] have characterized the performance of PbTe *p-n* junction, at 80 K, prepared on $BaF_2$. The values of $R_0A$ (0.23 – 31.8 $\Omega \cdot cm^2$) and $D^*$ (5.96×10$^8$ – 8.01×10$^{10}$ cm·Hz$^{1/2}$/W) they obtained, in different samples, are markedly lower than our results.

## 5. Summary

Thermally diffused and ion-implanted PbTe *p-n* junction diodes properties have been thoroughly investigated and characterized over a wide region of temperature and bias voltages. All the physical parameters of the diodes have been derived from an analysis of *C-V* and *I-V* measurements. The temperature dependence of the static dielectric constant of PbTe was determined, for the first time, from 10 K up to 300 K. The values of the measured and of the estimated parameters of these diodes demonstrate their high photodetector performance.

**ACKNOWLEDGMENT**

We wish to thank Dr V. Richter for ion implantation of PbTe crystals. This work has been supported partly by the Ministry of Science and Technology of Israel, in the framework of the China-Israel Joint Research Program on Advanced Materials 2005-6.

Table 1. Performance of PbTe detectors

| Sample | $T$ K | $R_0$ $\Omega$ | $R_0A_e$ Ohm·cm$^2$ | $j_0$ A/cm$^2$ | $R_0C$ sec | $D^*$ cm·Hz$^{1/2}$/W | $D_j^*$ cm·Hz$^{1/2}$/W |
|---|---|---|---|---|---|---|---|
| TDJ | 180 | 59 | 0.38 | 0.10 | $3.25\times10^{-7}$ | $1.0\times10^{10}$ | $1.1\times0^{10}$ |
| TDJ | 140 | 540 | 3.46 | $6.7\times10^{-3}$ | $3.09\times10^{-6}$ | $3.3\times10^{10}$ | $3.3\times10^{10}$ |
| TDJ | 80 | $1.32\times10^5$ | 845 | $8.8\times10^{-6}$ | $9.06\times10^{-4}$ | $1.2\times10^{11}$ | $7.1\times10^{11}$ |
| TDJ | 12.7 | $1.3\times10^7$ | $8.3\times10^4$ | $3.0\times10^{-11}$ | 0.086 | $1.2\times10^{11}$ | $1.7\times10^{13}$ |
| In-IJ | 180 | 19 | 0.12 | 2.2 | - | $5.6\times10^9$ | $5.6\times10^9$ |
| In-IJ | 140 | 72 | 0.46 | $5.3\times10^{-2}$ | - | $1.2\times10^{10}$ | $1.2\times10^{10}$ |
| In-IJ | 80 | $3.92\times10^4$ | 250 | $3.61\times10^{-5}$ | $1.82\times10^{-3}$ | $1.2\times10^{11}$ | $3.8\times10^{11}$ |
| In-IJ | 12.6 | $2.42\times10^6$ | $1.55\times10^4$ | $4.4\times10^{-10}$ | 0.17 | $1.2\times10^{11}$ | $7.6\times10^{12}$ |
| Zn-IJ | 180 | 14.6 | 0.09 | 0.7 | - | $4.8\times10^9$ | $4.8\times10^9$ |
| Zn-IJ | 140 | **103** | 0.66 | $3.4\times10^{-2}$ | $1.61\times10^{-6}$ | $1.5\times10^{10}$ | $1.5\times10^{10}$ |
| Zn-IJ | 80 | $1.24\times10^4$ | 79.36 | $9.0\times10^{-5}$ | $1.79\times10^{-4}$ | $1.1\times10^{11}$ | $2.16\times10^{11}$ |

**Figure captions**

Fig. 1. Schematic view of the *p-n* junction diode array.

(a) Thermally diffused junctions prepared by $In_3Te_3$ gas diffusion into the mesa-columns. (b) The planar array of implanted junctions, formed by In or Zn ion-implantation in the windows opened in the oxidized layer.

Fig. 2. The capacitance-voltage characteristics of TDJ at different temperatures.

(1) 150 K; (2) 110 K; (3) 80 K; (4) 60.1 K; (5) 42.2 K; (6), 25.6 K; (7) 12.8 K

Fig. 3. The capacitance-voltage characteristics of (a) In-IJ and (b) Zn-IJ at different temperatures.

(a) In-IJ: (1) 105 K; (2) 90 K; (3) 70.1 K; (4) 50.1 K; (5) 25.6 K

(b) Zn-IJ: (1) 100 K; (2) 80 K; (3) 60.1 K; (4) 45.2 K; (5) 26.6 K; (6) 12.6 K

Fig. 4. The temperature dependence of the static dielectric constant $\varepsilon(T)$ constructed by the fitting procedure.

The solid line is fitting; the double line is the high temperature "reference" line calculated from Eq. (12).

Fig. 5. The temperature dependences of the energy gap, $E_g$ and the barrier height, $V_0$.

(a) TDJ, (b) In-IJ, (c) Zn-IJ. 1 - $V_0$ determined by Eq. (2); 2 - $V_0$ determined by Eq. (9), i.e. measured directly.

Fig. 6. The doping gradients *a* for TDJ (1, left axis), In-IJ (2, right axis), and Zn-IJ (3, left axis). $a_1 = 2.01 \cdot 10^{21}$ cm$^{-4}$, $a_2 = 4.57 \cdot 10^{23}$ cm$^{-4}$, $a_3 = 5.73 \cdot 10^{21}$ cm$^{-4}$.

Fig. 7. The width of the *p-n* junction (left axis) and the dopant concentrations at the junction boundary (right axis) for (a) TDJ, (b) In-IJ, and (c) Zn-IJ.

Fig. 8. The hole, $F_p$ (1) and electron, $F_n$ (2) Fermi levels for (a) TDJ, (b) In-IJ and (c) Zn-IJ.

Fig. 9. The current-voltage characteristics of TDJ at different temperatures.

(a): (1) 24.7 K; (2) 40.2 K; (3) 60.1 K; (4) 80 K;

(b): (5) 100 K; (6) 120 K; (7) 140 K.

Fig.10. The fitting of current-voltage characteristics of TDJ.

Gray circles – experiment; black lines – fitting.

Fig. 11. Temperature dependence of the dynamical resistance at zero bias voltage.

(1) TDJ; (2) In-IJ; (3) Zn-IJ.

Fig. 12. (a) The ideality coefficient vs. temperature for: (1) TDJ; (2) In-IJ; (3) Zn-IJ.

(b) The saturation current density vs. temperature for: (1) TDJ; (2) In-IJ; (3) Zn-IJ.

Fig. 13. The current-voltage characteristics for (a) In-IJ and (b) Zn-IJ at different temperatures.

(a): (1) 140 K; (2) 122 K; (3) 80 K; (4) 60 K; (5) 14.7 K

(b): (1) 140 K; (2) 122 K; (3) 80 K; (4) 60 K; (5) 16.4 K

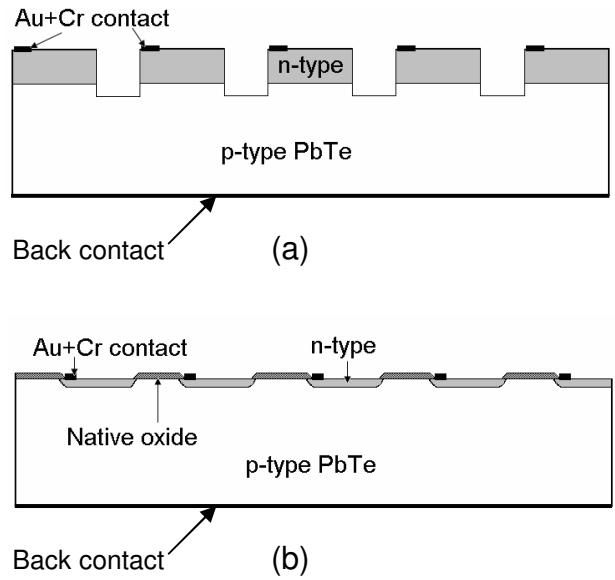

Fig. 1. Schematic view of the *p-n* junction diode array.

(a) Thermally diffused junctions prepared by In$_3$Te$_3$ gas diffusion into the mesa-columns. (b) The planar array of implanted junctions, formed by In or Zn ion-implantation in the windows opened in the oxidized layer.

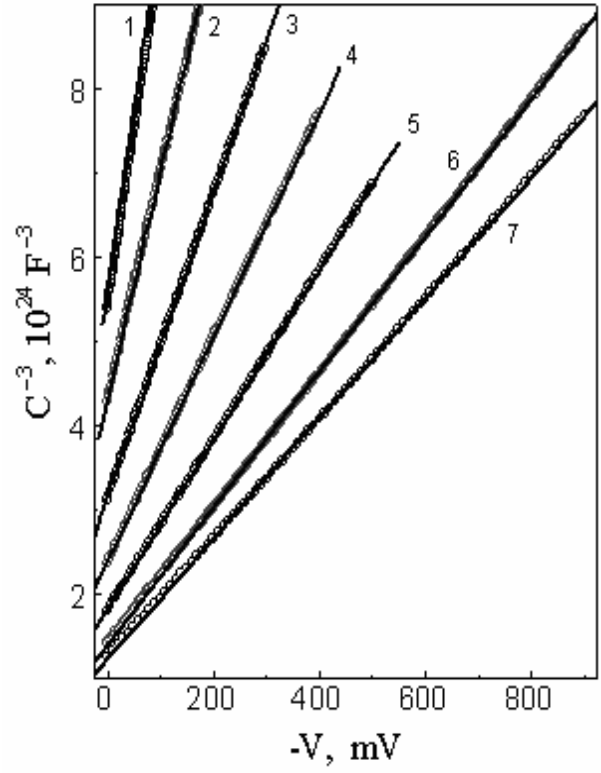

Fig. 2. The capacitance-voltage characteristics of TDJ at different temperatures.

(1) 150 K; (2) 110 K; (3) 80 K; (4) 60.1 K; (5) 42.2 K; (6), 25.6 K; (7) 12.8 K

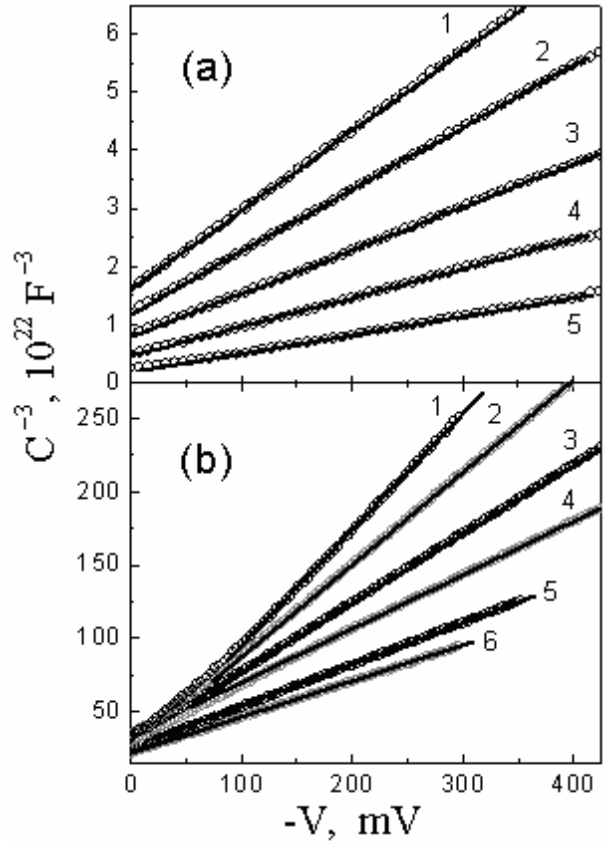

Fig. 3. The capacitance-voltage characteristics of (a) In-IJ and (b) Zn-IJ at different temperatures.

(a) In-IJ: (1) 105 K; (2) 90 K; (3) 70.1 K; (4) 50.1 K; (5) 25.6 K

(b) Zn-IJ: (1) 100 K; (2) 80 K; (3) 60.1 K; (4) 45.2 K; (5) 26.6 K; (6) 12.6 K

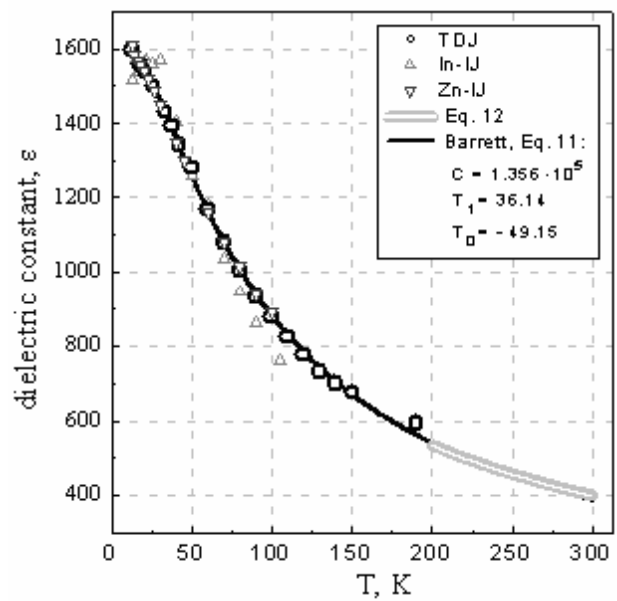

Fig. 4. The temperature dependence of the static dielectric constant $\varepsilon(T)$ constructed by the fitting procedure.

The solid line is fitting; the double line is the high temperature "reference" line calculated from Eq. (12).

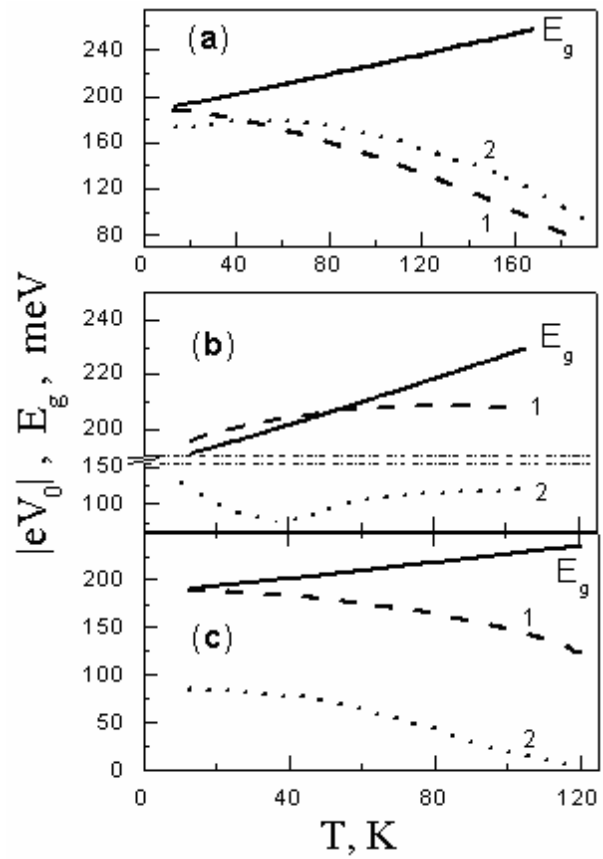

Fig. 5. The temperature dependences of the energy gap, $E_g$ and the barrier height, $V_0$. (a) TDJ, (b) In-IJ, (c) Zn-IJ. 1 - $V_0$ determined by Eq. (2); 2 - $V_0$ determined by Eq. (9), i.e. measured directly.

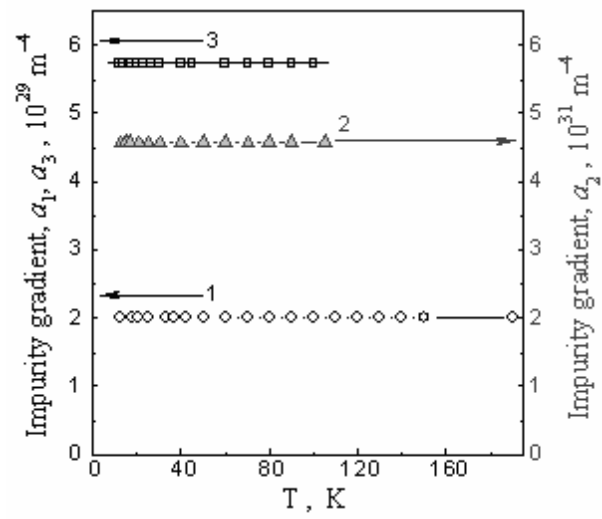

Fig. 6. The doping gradients $a$ for TDJ (1, left axis), In-IJ (2, right axis), and Zn-IJ (3, left axis). $a_1 = 2.01 \cdot 10^{21}$ cm$^{-4}$, $a_2 = 4.57 \cdot 10^{23}$ cm$^{-4}$, $a_3 = 5.73 \cdot 10^{21}$ cm$^{-4}$.

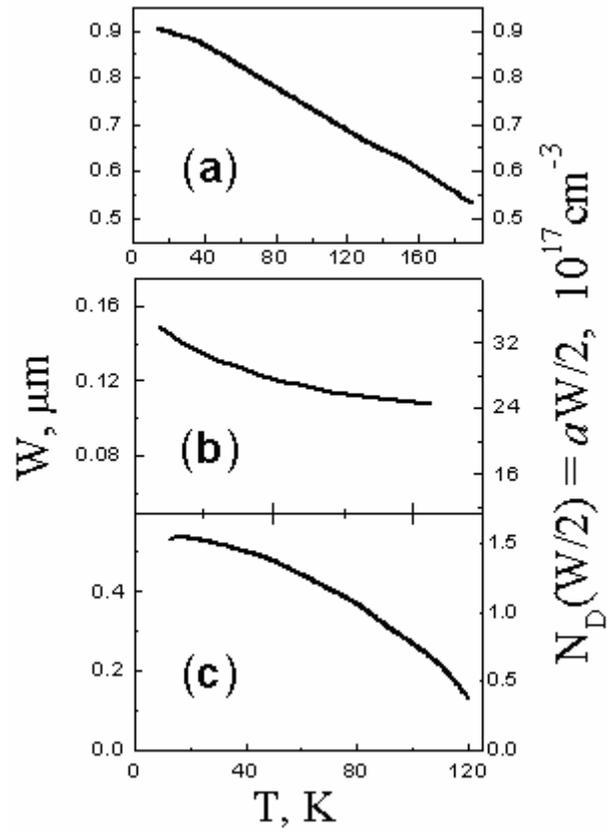

Fig. 7. The width of the *p-n* junction (left axis) and the dopant concentrations at the junction boundary (right axis) for (a) TDJ, (b) In-IJ, and (c) Zn-IJ.

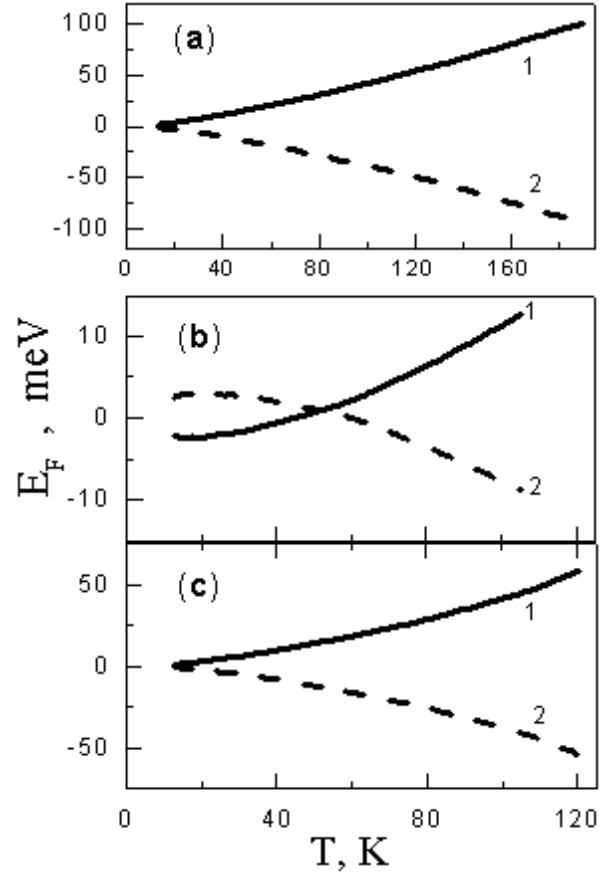

Fig. 8. The hole, $F_p$ (1) and electron, $F_n$ (2) Fermi levels for (a) TDJ, (b) In-IJ and (c) Zn-IJ.

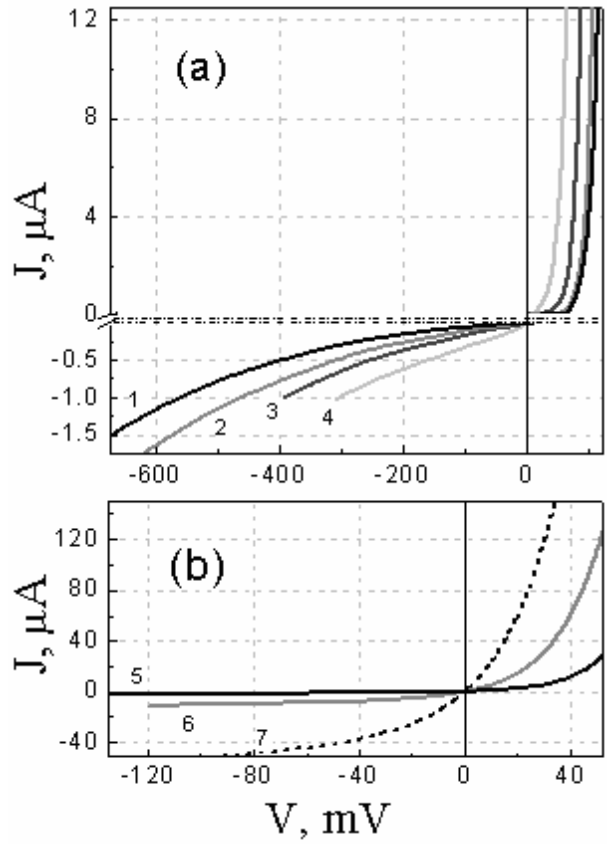

Fig. 9. The current-voltage characteristics of TDJ at different temperatures.

(a): (1) 24.7 K; (2) 40.2 K; (3) 60.1 K; (4) 80 K;

(b): (5) 100 K; (6) 120 K; (7) 140 K.

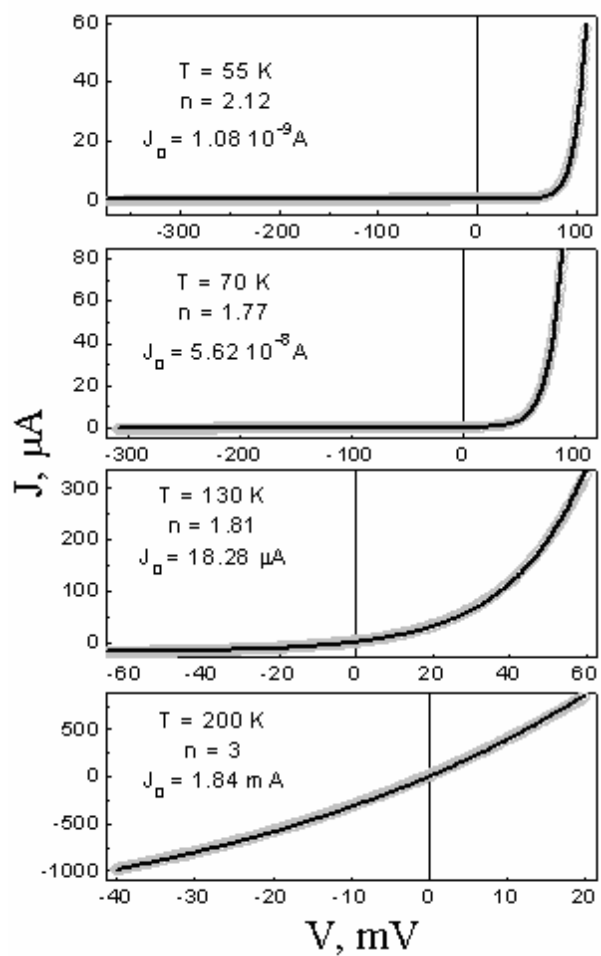

Fig.10. The fitting of current-voltage characteristics of TDJ.

Gray circles – experiment; black lines – fitting.

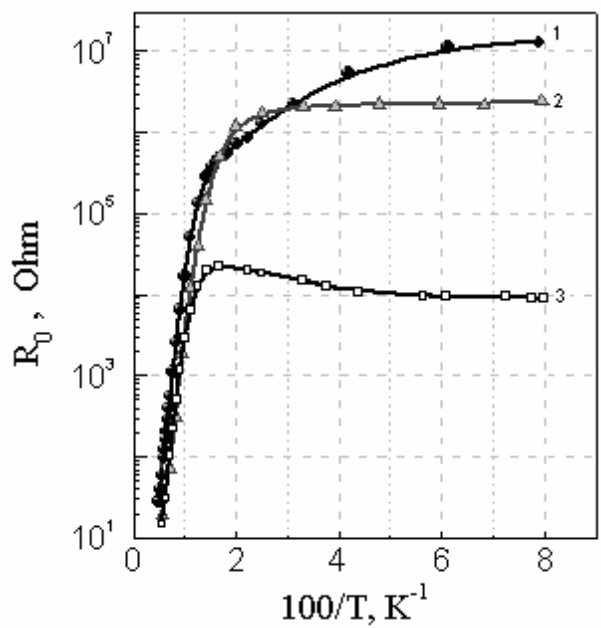

Fig. 11. Temperature dependence of the dynamical resistance at zero bias voltage.
(1) TDJ; (2) In-IJ; (3) Zn-IJ.

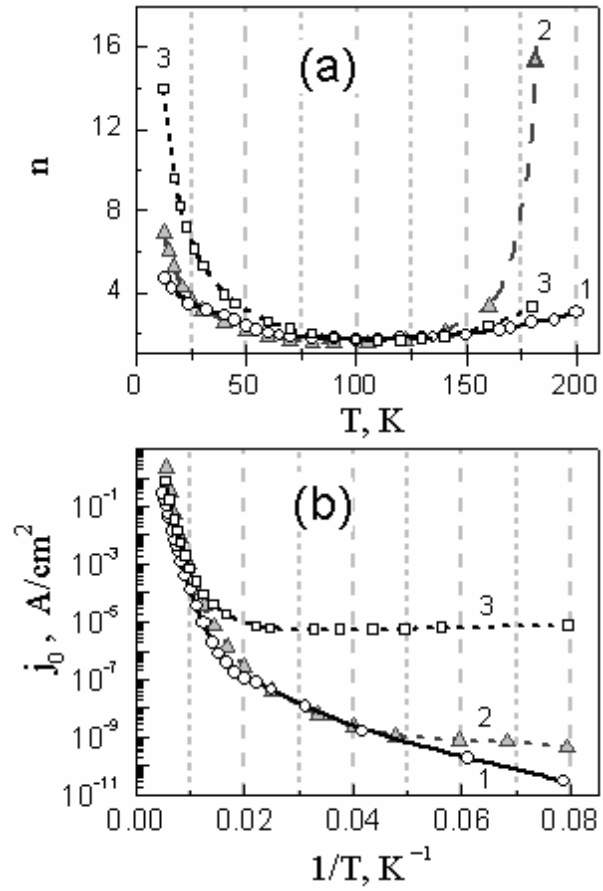

Fig. 12. (a) The ideality coefficient vs. temperature for: (1) TDJ; (2) In-IJ; (3) Zn-IJ.
(b) The saturation current density vs. temperature for: (1) TDJ; (2) In-IJ; (3) Zn-IJ.

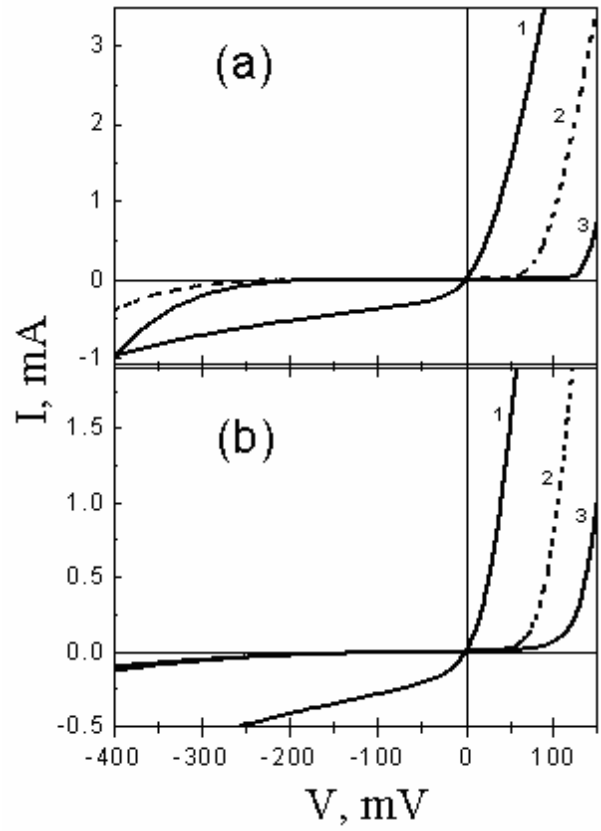

Fig. 13. The current-voltage characteristics for (a) In-IJ and (b) Zn-IJ at different temperatures.

(a): (1) 140 K; (2) 122 K; (3) 80 K; (4) 60 K; (5) 14.7 K

(b): (1) 140 K; (2) 122 K; (3) 80 K; (4) 60 K; (5) 16.4 K